\documentclass[acmsmall,screen]{acmart}
\usepackage{booktabs} % For formal tables
\usepackage{multirow}
\usepackage{subfigure}
\usepackage{array}
\usepackage{float}
\usepackage{hyperref}
\usepackage{setspace}
\usepackage{threeparttable}
\usepackage{supertabular}
\usepackage{bm}
\usepackage{amsmath}
\usepackage{amsthm}
\usepackage{mathrsfs}

\newcommand{\blue}{\textcolor{black}}
\AtBeginDocument{%
  \providecommand\BibTeX{{%
    \normalfont B\kern-0.5em{\scshape i\kern-0.25em b}\kern-0.8em\TeX}}}

\begin{document}

\title{An Intent Taxonomy of Legal Case Retrieval}

\author{Yunqiu Shao}
\email{shaoyq18@mails.tsinghua.edu.cn}
\affiliation{%
  \institution{Department of Computer Science and Technology, Quan Cheng Laboratory, Institute for Internet Judiciary, Tsinghua University}
  \city{Beijing}
  \country{China}
}

\author{Haitao Li}
\email{liht22@mails.tsinghua.edu.cn}
\affiliation{%
  \institution{Department of Computer Science and Technology, Quan Cheng Laboratory, Institute for Internet Judiciary, Tsinghua University}
  \city{Beijing}
  \country{China}
}

\author{Yueyue Wu}
\email{wuyueyue1600@gmail.com}
\affiliation{%
  \institution{Department of Computer Science and Technology, Quan Cheng Laboratory, Institute for Internet Judiciary, Tsinghua University}
  \city{Beijing}
  \country{China}
}

\author{Yiqun Liu}
\email{yiqunliu@tsinghua.edu.cn}
\authornote{Corresponding author}
\affiliation{%
  \institution{Department of Computer Science and Technology, Quan Cheng Laboratory, Institute for Internet Judiciary, Tsinghua University}
  \city{Beijing}
  \country{China}
}

\author{Qingyao Ai}
\email{aiqy@tsinghua.edu.cn}
\affiliation{%
  \institution{Department of Computer Science and Technology, Quan Cheng Laboratory, Institute for Internet Judiciary, Tsinghua University}
  \city{Beijing}
  \country{China}
}

\author{Jiaxin Mao}
\email{maojiaxin@gmail.com}
\affiliation{%
  \institution{Gaoling School of Artificial Intelligence, Renmin University of China}
  \city{Beijing}
  \country{China}
}

\author{Yixiao Ma}
\email{ma-yx16@tsinghua.org.cn}
\affiliation{%
  \institution{Department of Computer Science and Technology, Quan Cheng Laboratory, Institute for Internet Judiciary, Tsinghua University}
  \city{Beijing}
  \country{China}
}

\author{Shaoping Ma}
\email{msp@tsinghua.edu.cn}
\affiliation{%
  \institution{Department of Computer Science and Technology, Quan Cheng Laboratory, Institute for Internet Judiciary, Tsinghua University}
  \city{Beijing}
  \country{China}
}

\renewcommand{\shortauthors}{Shao, et al.}

\begin{abstract}
  Legal case retrieval is a special Information Retrieval~(IR) task focusing on legal case documents. Depending on the downstream tasks of the retrieved case documents, users' information needs in legal case retrieval could be significantly different from those in Web search and traditional ad-hoc retrieval tasks. While there are several studies that retrieve legal cases based on text similarity, the underlying search intents of legal retrieval users, as shown in this paper, are more complicated than that yet mostly unexplored. To this end, we present a novel hierarchical intent taxonomy of legal case retrieval. It consists of five intent types categorized by three criteria, i.e., search for \textit{Particular Case(s)}, \textit{Characterization}, \textit{Penalty}, \textit{Procedure}, and \textit{Interest}. The taxonomy was constructed transparently and evaluated extensively through interviews, editorial user studies, and query log analysis. Through a laboratory user study, we reveal significant differences in user behavior and satisfaction under different search intents in legal case retrieval. Furthermore, we apply the proposed taxonomy to various downstream legal retrieval tasks, e.g., result ranking and satisfaction prediction, and demonstrate its effectiveness. Our work provides important insights into the understanding of user intents in legal case retrieval and potentially leads to better retrieval techniques in the legal domain, such as intent-aware ranking strategies and evaluation methodologies. 
\end{abstract}

% Our work provides insight into user intents in legal case retrieval. It facilitates legal case retrieval systems to better satisfy user information needs in practice, such as developing intent-aware ranking strategies and suitable evaluation methodologies. 





\maketitle

\section{Introduction}
%% legal case search
% Legal case documents are primary legal materials in various law systems. 
% As a specialized Information Retrieval~(IR) task, legal case retrieval aims to search for legal cases relevant to the user's matter or problem. 
\blue{
Legal case retrieval plays a crucial role in modern legal systems~\cite{laaa008}. In countries that follow case law system, judges rely on previous judgments of relevant cases to reach a final decision~\cite{shulayeva2017recognizing}. In countries with statutory law system, extensive examination of pertinent cases is conducted when presenting a case to the court to prevent erroneous judgments~\cite{hamann2019german}.} With the rapid growth of digitalized case documents, legal case retrieval has attracted increasing attention in both IR and legal communities. Existing research efforts~\cite{rabelo2019summary, xiao2019cail2019, bhattacharya2019overview, ma2021lecard} on legal case retrieval mostly focus on estimating and measuring case similarity. For instance, the CAIL2019-SCM~\cite{xiao2019cail2019} task focuses on comparing the similarity between cases in each case triplet. The COLIEE~\cite{rabelo2019summary} and AILA~\cite{bhattacharya2019overview} benchmarks are designed to evaluate retrieval systems' ability in identifying supporting cases regarding a query case. However, as shown in this paper, the application scenario of legal case retrieval is broader and more complicated than similar case matching. Without knowing the actual needs of legal search users, it is difficult to develop a legal case retrieval system that is effective and reliable.
% In particular, legal case retrieval is usually conducted to obtain different types of fine-grained information, such as opinions, decisions, and case details, that may be contained in the relevant case documents. The underlying user intents in legal case retrieval might be diverse and more complex than known item search~\cite{rabelo2019summary}. 
% which still lack an in-depth understanding. 
% Under the framework of measuring the relationship between cases, a variety of retrieval models~\cite{shao2020bert,westermann2020paragraph,mandal2021unsupervised, yu2022explainable} have been proposed.

%% intent taxonomy, the effects of user intent
In fact, how to understand and model search intents has been a fundamental research question for IR research~\cite{broder2002taxonomy,rose2004understanding,kofler2016user,su2018user,xie2018people,cambazoglu2021intent}. For example, the popular taxonomy of Web search intents proposed by Broder~\cite{broder2002taxonomy}~(i.e., navigational, informational, and transactional) has been widely used in the interpretation of user behavior and the design of retrieval models. It has profound implications for subsequent researches in both algorithm and evaluation design. Under different search intents, the user's expected results, search behavior, and satisfaction perception can be different significantly~\cite{jiang2014searching,chapelle2011intent,wu2019influence,bolotova2022non}. Therefore, methodologies in search systems, including relevance criteria, ranking strategies, and evaluation metrics, must be adapted to different search intents accordingly~\cite{chapelle2011intent,white2013enhancing,kharitonov2013intent,mehrotra2019jointly,bolotova2022non}.

%% the character of legal case search.
The legal case retrieval scenario differs from general Web search remarkably. Specifically, the users of legal case retrieval are mainly legal practitioners with professional knowledge. The retrieved results are primarily authoritative case documents containing rich legal knowledge rather than web content with different quality levels. Instead of Web search engines, professional legal search tools~(e.g., \textit{WestLaw}, \textit{LexisNexis}) are preferred~\cite{arewa2006open,barkan2015fundamentals}. Recent research~\cite{shao2021investigating} has also pointed out that users' search behavior in legal case retrieval differs significantly from that in Web search. Therefore, domain-specific characteristics should also be considered regarding the search intents in legal case retrieval. 
However, to our best knowledge, there still lacks a well-defined taxonomy of search intents in legal case retrieval. Existing taxonomies in legal information systems are mainly designed based on legal issues and topics, such as the ``Key Number System''~\cite{sprowl1975westlaw}, while the underlying user intents \blue{are not sufficiently studied}. 

Toward a legal case retrieval system that can better satisfy diverse user information needs, this paper takes an in-depth investigation into user intents. Specifically, our research questions are:
% and their effects in legal case retrieval. Specifically, our research questions are:
\begin{itemize}
    \item \textbf{RQ1:} \textit{What are the types of user intent in legal case retrieval?}
    \item \textbf{RQ2:} \textit{How \blue{does} user search behavior change with search intents in legal case retrieval?}
    \item \textbf{RQ3:} \textit{What are the differences in perception and measurement of user satisfaction under different search intents? }
    \item \textbf{RQ4:} \textit{How can the taxonomy benefit downstream tasks in legal case retrieval?}
\end{itemize}

Regarding \textbf{RQ1}, we proposed a hierarchical intent taxonomy of legal case retrieval, which integrates IR and legal classification theory. To construct the taxonomy, we inspected the user surveys collected from legal practitioners and real-life queries issued to the commercial legal case retrieval engine. The taxonomy was further verified through interviews and editorial user studies. We also present the distributions of intents in legal case retrieval. To the best of our knowledge, it is the first intent taxonomy designed for legal case retrieval. 

% With the established taxonomy, we move on to investigate what differences the intent taxonomy would make to legal case retrieval. In particular, we focus on answering the following research questions:

To address the above RQs, we conducted a laboratory user study with participants majoring in law. Rich behavioral data were logged to inspect the search process under different search intents. Besides, we collected explicit user feedback, such as user satisfaction and clicked reasons, to understand how users' perceptions of satisfaction change with different search intents. We also shed light on evaluating legal case retrieval across different search intents based on online metrics. Furthermore, we applied the intent categories to different downstream IR tasks, such as satisfaction prediction and result ranking. Our results reveal the significant impacts of the intent taxonomy on legal case retrieval. 

% Our results reveal significant differences in user behavior and satisfaction under different intent categories in legal case retrieval. 

% , including queries, click-through, hovers, and timestamps

% Given the above findings, our further question is \textit{``Can we identify the search intent timely to facilitate the legal case retrieval system providing proper results accordingly?''} As an attempt to address it, we built an intent classifier based on the input query and evaluated its effectiveness on real-life query logs. 

To summarize, our key contributions are as follows:
\begin{itemize}
    \item We propose a novel intent taxonomy of legal case retrieval. \blue{The taxonomy has five intent categories, i.e., search for \textit{Particular Case(s)}, \textit{Characterization}, \textit{Penalty}, \textit{Procedure}, and \textit{Interest}}. To our best knowledge, it is the first taxonomy that categorizes users' search intents in legal case retrieval.
    \item The taxonomy was constructed and evaluated extensively using multiple resources, such as interviews, editorial user studies, log analysis, etc. We provide the formal procedure of taxonomy creation. Moreover, we reveal the distributions of different search intents in the realistic search scenario of legal case retrieval. 
    \item We collected a behavioral dataset with user satisfaction feedback under the proposed intent taxonomy via a controlled laboratory user study. We show significant differences in multiple search behavior patterns with different search intents. The dataset will be open after acceptance. 
    \item Regarding user satisfaction, we illustrate significant differences in users' perceptions of satisfaction and the different influential factors under different search intents. 
    \item We applied the intent taxonomy to common downstream tasks, including satisfaction prediction and result ranking. Experimental results demonstrate its benefits and effectiveness.   
\end{itemize}
% \item To aid the legal case retrieval system respond to the underlying user intent in time, we build an intent predictor with the query content. Experiments on real-life query logs demonstrate its effectiveness.
% The taxonomy was constructed in a transparent way and extensively verified via interviews and editorial user studies. We also present the distributions of intents in legal case retrieval. 
% We conducted a controlled laboratory user study given different user intents in legal case retrieval.
% , including query formulation, page-visiting, hovering, clicking, and dwell time patterns
 % by comparing online metrics with user satisfaction
 % Taking user satisfaction as the ``golden standard'' of search evaluation, we also find the impacts on online evaluation. 

The rest of this paper is organized as follows. Section~\ref{sec: related work} reviews the related work. Section~\ref{sec:overview} provides an overview of the proposed intent taxonomy in legal case retrieval. Section~\ref{sec: construction} describes the construction procedure of the taxonomy. Section~\ref{sec:behavior_satis} focuses on answering \textbf{RQ2} and \textbf{RQ3}, which introduces the user study settings and findings in user behavior and satisfaction. Section~\ref{sec:application} presents the taxonomy's applications to satisfaction prediction and ranking tasks regarding \textbf{RQ4}. Section~\ref{sec: discussion} discusses the main findings in this paper and significant implications. Finally, Section~\ref{sec: conclusion} discusses the conclusions, potential limitations, and future work directions. 

\section{Related Work}
\label{sec: related work}

\subsection{Search Intent Taxonomy}
It is a fundamental task for IR systems to understand users' search intents to satisfy diverse types of information needs. In  Web search, Broder~\cite{broder2002taxonomy} proposed a widely adopted taxonomy using a user survey and analysis of query logs in AltaVista. According to the ``need behind query'', the taxonomy classified web queries into three categories, navigational, informational, and transactional. Based on this taxonomy, Rose and Levinson~\cite{rose2004understanding} proposed a more precise classification framework from the perspective of understanding why users are searching. Recently, Cambazoglu et al.~\cite{cambazoglu2021intent} built a new multi-faceted intent taxonomy for questions asked in Web search engines based on 1,000 real-life issued questions, which was more fine-grained but less ambiguous for human assessors. Bolotova et al.~\cite{bolotova2022non} presented a comprehensive taxonomy of the current non-factoid question-answering task and they also pointed out that the challenging categories were poorly represented in the existing datasets.  
Besides web search, search intents have also been investigated in some specific search scenarios, such as multi-media search~\cite{kofler2016user}, image search~\cite{lux2010classification,xie2018people}, product search~\cite{sondhi2018taxonomy,su2018user}, medical search~\cite{wang2022recognizing}, etc. 
Given different search intents, some research indicated that user search behavior would also be different~\cite{su2018user,xie2018people,wu2019influence,jiang2014searching}. Meanwhile, with a good understanding of underlying search intents and related effects, search engines could be further improved in various aspects, such as diversity search~\cite{chapelle2011intent}, personalized search~\cite{white2013enhancing}, query suggestion~\cite{kharitonov2013intent}, result ranking~\cite{glater2017intent}, satisfaction prediction~\cite{mehrotra2019jointly}, and system evaluation~\cite{zhang2018well,bolotova2022non}. 

Meanwhile, taxonomies in the legal field are almost centered on objective legal knowledge, such as classifying law systems~\cite{siems2016varieties}, rules of law~\cite{sherwin2009legal}, and legal issues~\cite{sprowl1975westlaw}. For example, the well-known ``Key Number System'' in Westlaw~\cite{sprowl1975westlaw} is a kind of taxonomy that organize cases by legal issues and topics. 
However, user search intents were not included in these taxonomies. To the best of knowledge, there is no systematic modeling of search intents in legal case retrieval.
% lacking a systematic modeling. 

\subsection{Legal Case Retrieval}
Legal case retrieval is a specialized IR task that aims to search for relevant legal cases given the matter at hand. \blue{Compared to web search, legal case retrieval has unique challenges.On the one hand, legal documents are often much longer and use domain-specific terminology than web document. On the other hand, the definition of relevance for legal case retrieval goes beyond simple semantic similarity.} It is a crucial task in legal practice and has drawn active research efforts in both legal and IR communities. In the earlier decades, extensive expert efforts were invested in organizing legal knowledge and developing the professional legal information system. \blue{For example, Moens ~\cite{moens2001innovative} identified some form of concept based retrieval, containing three models: Boolean retrieval, vector space retrieval, and probabilistic retrieval. Klein et al. ~\cite{klein2006thesaurus} outlined ontological-based approaches for retrieving similar cases to a seed case.} In recent years, with the rapid increase of digitalized case documents and the development of NLP techniques, research efforts have been put into developing automatic retrieval models that can identify relevant cases given a query case. Several benchmarks have been constructed for this task, such as COLIEE~\cite{rabelo2019summary}, AILA~\cite{bhattacharya2019overview}, CAIL2019-SCM~\cite{xiao2019cail2019}, and LeCaRD~\cite{ma2021lecard}, where the \blue{core concern} is to measure the semantic relationship~(e.g., similarity) between cases. For instance, in CAIL2019-SCM~\cite{xiao2019cail2019}, the task is to detect which two cases are more similar in each triplet. Based on these benchmarks, a variety of case retrieval models~\cite{shao2020bert, westermann2020paragraph, mandal2021unsupervised, yu2022explainable, li2023sailer, li2023} have been proposed, such as measuring case relevance via automatic summarization~\cite{rossi2019legal,tran2019building}, paragraph semantic modeling~\cite{shao2020bert,westermann2020paragraph}, rationale matching~\cite{yu2022explainable}, and so on. \blue{LOCKE et al.~\cite{locke2022case} summarize the methods of case law retrieval in the past 30 years and point out that the future of case law retrieval is based on natural language. For example, Savelka et al. utilize pre-trained language models to discover explanatory sentences for legal cases~\cite{savelka-ashley-2021-discovering-explanatory,vsavelka2022legal,savelka2019improving}. Li et al. utilize the structure of legal cases to design new pre-training objectives, which yielded state-of-the-art results on legal case retrieval~\cite{li2023sailer,li2023thuircoliee}.}
Beyond developing semantic case-matching models, some recent works pay attention to user-system interactions in legal case retrieval. In particular, Shao et al.~\cite{shao2021investigating} conducted a user study to investigate user behavior in legal case retrieval and illustrated significant differences from the web search scenario, e.g., the exploratory property. Given the task complexity in legal case search, Liu et al.~\cite{liu2021conversational} attempted to apply the conversational search paradigm, which might facilitate the users expressing their information needs better, according to their user study. It is noteworthy that the tasks in these studies were still designed based on the classification of objective legal knowledge, such as the rules of law~\cite{shao2020bert} or the legal issues~\cite{liu2021conversational}. However, as far as we know, users' underlying search intents in legal case retrieval have not been investigated systematically nor considered in existing studies, such as retrieval model development or user study design.

% Towards understanding the practical search process in legal case retrieval, a recent study~\cite{shao2021investigating} investigated user behavior via a user study and illustrated significant differences from the web search scenario. However, as far as we know, the diverse underlying search intents have not been considered in existing studies, either in retrieval model development or user study design. 

% \subsection{User Behavior and Satisfaction}

% However, regarding information needs of legal practitioners, though there were some empirical studies~\cite{otike1999information,haruna2001information} working on the general information seeking process, they still lacked a systematic taxonomy, especially in the legal case retrieval. 

% \subsection{Legal Case Retrieval}
% Taxonomy of law systems~\cite{siems2016varieties}.
% Westlaw~\cite{sprowl1975westlaw}
% Legal Taxonomy, classify rules of law. ~\cite{sherwin2009legal}. 
% information needs and habits of lawyers ~\cite{otike1999information}. 

\section{Taxonomy Overview}

\label{sec:overview}

\blue{In this paper, we focus on the hierarchical intent taxonomy for legal case retrieval. We attempt to construct and validate the intent taxonomy in the Chinese legal system.
Specifically, the judicial process in China consists of four steps: prosecution, acceptance, preparation for trial, and trial. After the trial, the court will write a legal case document, consisting of basic information such as facts, reasoning, and the judgment. These legal case documents can be of great help to users. Legal practitioners can retrieve similar cases to help make decisions, and ordinary people can learn more about the law from legal cases. Our taxonomy is oriented to real retrieval scenarios. In real scenarios, users retrieve similar cases in order to solve the matter at hand. This means that the form of the question can be varied in a particular intent. In short, the corpus of the legal case retrieval is a large number of legal documents, while the query may be a case, a sentence or some keywords. }

\begin{figure}
    \centering
    \includegraphics[width=1.0\columnwidth]{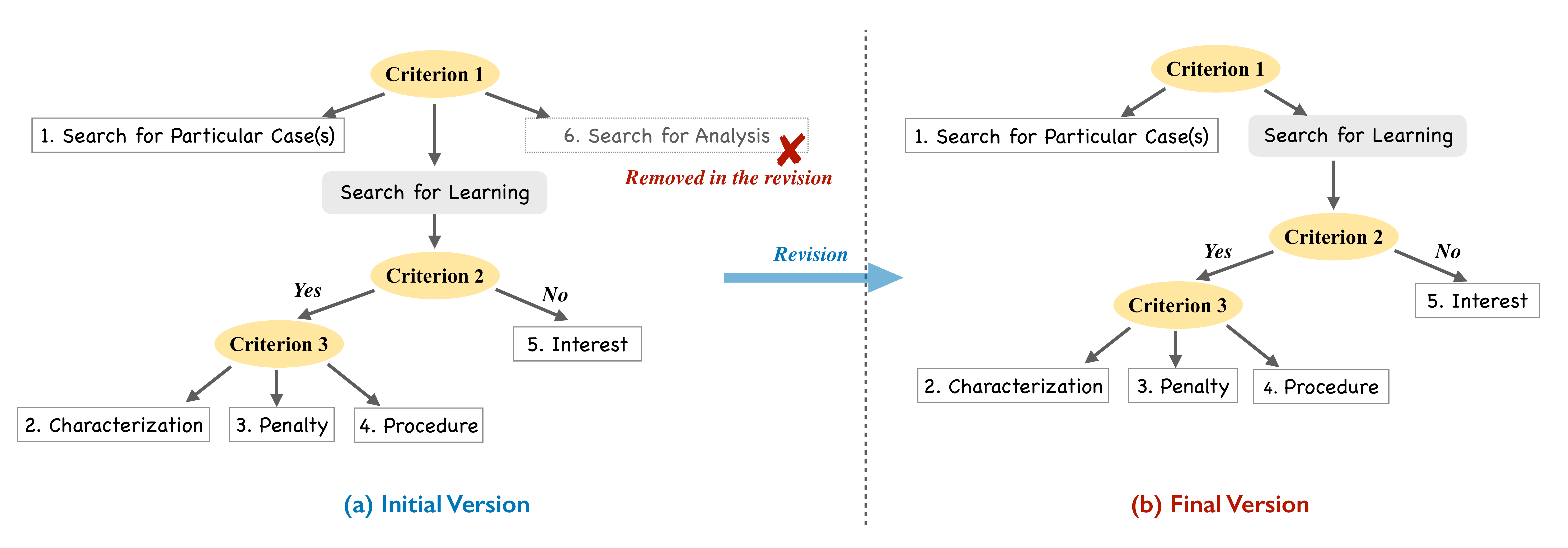}
    \caption{Illustrations of the proposed taxonomy structures. The (a) is the initial version of taxonomy and the (b) is the final version. }
    \label{fig:taxonomy}
\end{figure}

\blue{
There is no doubt that there is search intent when users enter a query in the legal case retrieval, i.e., a certain aspect of the case is expected to be retrieved. This intent may be explicit or implicit. A good case  retrieval system should facilitate the users to express different intents and return appropriate cases. Therefore, to better guide the design of the case retrieval system, we propose a novel hierarchical intent taxonomy of legal case retrieval, which integrates IR and legal classification theory. This section gives an overview of the final intent taxonomy, as shown in Table~\ref{tab:taxonomy} and Figure~\ref{fig:taxonomy}.} Figure~\ref{fig:taxonomy}~(the right one) illustrates the general framework of the proposed intent taxonomy. Specifically, the following three criteria are utilized to categorize user intents successively.

\begin{table*}\small
    \centering
    \caption{The proposed intent categories in legal case retrieval and examples. The \textit{Analysis} category was removed in the taxonomy revision process. \blue{It's worth noting that the Example is not a query that the user enters into the search engine, but rather a problem that they are currently attempting to address.}} 
    \label{tab:taxonomy}
    \begin{tabular}{p{0.25\columnwidth}p{0.4\columnwidth}p{0.3\columnwidth}}
    \toprule
    \textbf{Category} & \textbf{Description} & \textbf{Example}\\
    \midrule
    1. Particular Case(s) (PC) & Search for some particular case document(s), e.g., the judgment documents of a specific case, the parties’ previous convictions or lawsuits. & What are the lawsuits that Company A has involved? \\
    \midrule
    2. Characterization (Ch) & Search for learning about conviction or law application under the substantive law regarding the current issue. \blue{With this intention, users focus on the characteristics of different aspects of the underlying facts, such as different case types, causes, regions, statutes etc.} & 
    Whether trapping loans is constituted fraud?  
    
    \blue{Whether the claim is based on product liability or consumer fraud liability?}\\
    \midrule
    3. Penalty (Pe) & Search for learning about sentencing or penalty range regarding the current issue. \blue{Under this intent, users focus on the violations and corresponding penalties involved in the case, such as criminal penalties (imprisonment, probation), civil damages (economic damages, moral injury), administrative penalties (fines, revocation of business licenses), etc.} & What is the punishment for embezzling \$100,000 in XX? 
    
    \blue{Whether the request for the return of \$7,700 and punitive damages can be supported} \\
    \midrule
    4. Procedure (Pr) & Search for learning about procedure issues related to the procedural law, \blue{i.e., litigation procedures, appeal procedures, evidence collection procedures, and enforcement procedures.} & What procedure should be followed if an undergoing civil case involves a criminal offense? 
    
    \blue{Whether acts committed in 2017 are time-barred?}\\
    \midrule
    5. Interest (In) & Have no specific legal issue to solve but search for learning some related information to satisfy the individual interest. & Johnny Depp v. Amber Heard; What are the recent cases that apply the XX rule?\\
    \midrule
    \textit{6. Analysis} & \textit{Search for writing some analytical reports, e.g.,similar case search report, statistical survey on a specific charge.} & \textit{Writing a similar case search report regarding the XX case; An empirical study of corruption based on over 200 judgments.} \\
    \bottomrule
    \end{tabular}
\end{table*}

% We propose a novel hierarchical intent taxonomy of legal case retrieval regarding \textbf{RQ1}, which was established and verified in an iterative way. In this section, we first give an overview of the final intent taxonomy~(as shown in Table~\ref{tab:taxonomy} and Figure~\ref{fig:taxonomy}), and then describe the iterative creation procedure~(Figure~\ref{fig:procedure}).
% \subsection{Taxonomy Overview}

\begin{itemize}
    \item \textbf{Criterion~1} What is the purpose of legal case retrieval?
    \item \textbf{Criterion~2} Is the search driven by a clear objective or not?
    \item \textbf{Criterion~3} What kind of legal problem does the objective belong to? 
\end{itemize}
The first criterion~(\textbf{Criterion 1}) asks what users are searching for. According to this criterion, intents can be generally classified into two groups, search for \textit{Particular Case(s)}~(PC) and search for \textit{Learning}~(Le) from the cases. The \textit{PC} intent can be compared to a combination of navigational and transactional needs in Web search~\cite{broder2002taxonomy} or a known-item search in product search~\cite{su2018user}. Meanwhile, the \textit{Learning} category is somewhat similar to an informational need. Furthermore, the category \textit{Learning} involves multiple situations in legal case retrieval, and thus we apply the second criterion~(\textbf{Criterion 2}), concerning whether there is a clear objective to learn~\cite{rose2004understanding,xie2018people}. If not, we consider that the search session is to satisfy some individual interest, such as curiosity or gossip triggered by social news. Otherwise, we apply the third criterion (\textbf{Criterion 3}) to categorize the specific objective based on the general classification of law~\cite{williams1982learning}, i.e., substantive law\footnote{\url{https://en.wikipedia.org/wiki/Substantive_law}} or procedural law\footnote{\url{https://en.wikipedia.org/wiki/Procedural_law}}. Specifically, substantive law refers to the set of laws that governs how members of a society are to behave. In contrast, procedural law~(also referred to as adjective law) comprises the rules of procedures for making, administering, and enforcing substantive law. Note that this classification exists in different law systems, in other words, is generally applicable across law systems. 
Based on \textbf{Criterion 3}, we group search intents into three categories, \textit{Characterization}~(Ch), \textit{Penalty}~(Pe), and \textit{Procedure}~(Pr). The \textit{Procedure} intent is about issues under the procedural law. The \textit{Characterization} and \textit{Penalty} are classified based on two primary types of issues under the substantive law, such as crimes and punishments. 
Table~\ref{tab:taxonomy} provides the descriptions and examples of each intent category. \blue{It is worth noting that the example in table~\ref{tab:taxonomy} is not a query entered by the user but rather the intent. Users can construct various forms of queries to realize their search intent according to their own customs. For example, when a lawyer is dealing with a legal case, he would like to know the possible penalties for the defendant's behavior. Then he can input several keywords or this case into the search engine.}
% Among them, the \textit{Ch} and \textit{Pe} intents are classified based on two primary legal issues~(\red{meaning of legal issue}) under the substantive law, while the \textit{Pr} intent is about issues under the procedural law. 
% \footnote{\url{https://en.wikipedia.org/wiki/Procedural_law}}

Given different search intents, the results that users want to retrieve could have different properties. According to \textbf{Criterion 1}, users would have strong needs for both precision and recall under the \textit{PC} intent. The potential relevant cases might be definite and of a limited size, correspondingly. On the contrary, the relevant cases under \textit{Learning} could be broader. Furthermore, according to \textbf{Criterion 2}, one could expect that search under the \textit{Interest} intent would have relatively lower requirements on precision and recall compared to the others. Among the three types categorized by \textbf{Criterion 3}, \textit{Characterization} and \textit{Penalty} are based on the substantive law, while the \textit{Procedure} focuses on the issues under the procedural law. Therefore, the relevance criteria under the \textit{Procedure} might differ from those used in \textit{Characterization} and \textit{Penalty}. Meanwhile, comparing \textit{Penalty} with \textit{Characterization}, the information need under \textit{Penalty} would be more specific and precise, and thus precision would be more emphasized than under \textit{Characterization}. 

% Furthermore, comparing between \textit{Characterization} and \textit{Penalty}, the information need under \textit{Penalty} would be more specific and precise, and therefore precision would be more emphasized. Meanwhile, the \textit{Procedure} focuses on the issues under the procedural law, independent of the substantive law that \textit{Characterization} and \textit{Penalty} are based on, and thus its relevance criteria might also differ.
% Therefore, the scope of relevant cases regarding the \textit{Pr} might differ from those of \textit{Ch} and \textit{Pe}. 

In summary, we construct an intent taxonomy based on three criteria. The taxonomy consists of five intent categories, i.e., Search for \textit{Particular Cases}~(PC), \textit{Characterization}~(Ch), \textit{Penalty}~(Pe), and \textit{Procedure}~(Pr), and \textit{Interest}~(In), and the detailed construction process is described in the next section. \blue{It is worth noting that this legal classification theory mainly takes the Chinese legal system into consideration. We primarily verify the rationality of the taxonomy under the Chinese legal system. We believe that it can contribute to the legal community and inspire the development of taxonomies for different legal systems.}

\section{Construction Procedure}
\label{sec: construction}
\begin{figure}
\centering
\includegraphics[width=1.0\columnwidth]{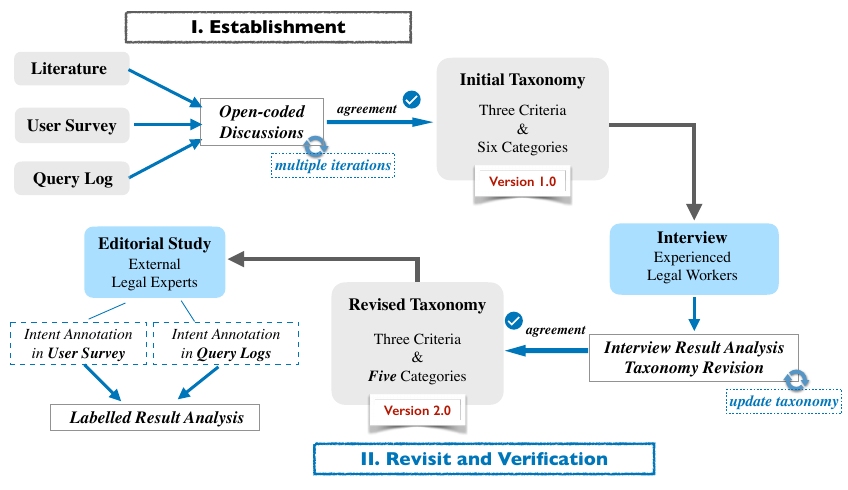}
\caption{The procedure of taxonomy creation, which consists of two stages (i.e., ``I. establishment'' and (2) 
``II. revisit and verification') in general. The second stage includes two parts, (1) semi-structured interviews and (2) an editorial user study. }
\label{fig:procedure}
\end{figure}

The taxonomy was constructed and evaluated extensively in an iterative way. In this section, we describe the creation procedure as illustrated in Figure~\ref{fig:procedure}. Generally, the procedure can be divided into two stages, \textit{I. establishment}, and \textit{II. revisit and verification}. 
In the establishment stage, we propose an initial version of the taxonomy. Then, we move on to verifying the taxonomy through different methods and revising it accordingly before settling down the final version. 

\subsection{Establishment}
\label{subsec:establish}
To construct the initial intent taxonomy, we exploited three resources, including the literature, user survey, and sampled query logs in the \textit{establishment} stage. We studied the literature on taxonomy in information retrieval and classification theory in the legal domain. In addition, we inspected users' real-life information needs in legal case retrieval through a user survey and query logs.

\paragraph{User Survey.} We designed an online survey to collect users' recent legal case retrieval experiences. Besides basic demographic questions, each participant was asked to answer the following three open questions according to her last time of legal case retrieval:
\begin{enumerate}
    \item What is the context that triggers this search? (e.g., working-on case, undergoing research topic, others' consultation, social news or hotpots, etc.)
    \item What is the detailed task of this search? \blue{(Specific queries and query intents, e.g.,client's litigation situation, equity betting cases, etc.)}
    \item What search engine(s) did you utilize in this search?
\end{enumerate}

\blue{The survey was spread via social media platforms, such as WeChat, etc.} Since the target users of legal case retrieval are mainly legal practitioners, we only collected responses from participants engaged in law-related occupations and paid each participant about \$1 for her serious response. We received responses from 116 participants and kept 110 after filtering the answers that were too vague or unrelated to legal case retrieval. The participants were from various legal-related occupations, including lawyers, staff of corporate legal affairs, prosecutors, judges and court staff, and legal researchers. \blue{Table ~\ref{Occupation} shows the occupational distribution of the participants. The user survey helped us gain a deeper understanding of the diverse search intents and tasks performed by legal professionals in their daily work.}

\begin{table}[t]
\centering
\caption{Occupational distribution of user survey participants. Court, Procuratorate, and Corporate represent the staff of each of them.}
\begin{tabular}{ccccccc}
\hline
Occupation & Court  & Procuratorate & Lawyers & Corporate & Legal Researcher & Other   \\ \hline
Number     & 10     & 17            & 31      & 18        & 14               & 20      \\ \hline
Ratio      & 9.09\% & 15.45\%       & 28.18\% & 16.36\%   & 12.73\%          & 18.18\% \\ \hline
\end{tabular}
\label{Occupation}
\end{table}

\paragraph{Query Logs.} We sampled 600 search sessions from a commercial legal case search engine\footnote{\url{https://ydzk.chineselaw.com/case}}. \blue{They were sampled from 7-day query logs in August 2021, involving 516 users.} Similar to previous studies~\cite{xie2018people}, we used 30 minutes as the window for splitting sessions. Furthermore, we excluded the sessions with less than one query term, which might be too vague to identify the search intents~\cite{shao2021investigating}. \blue{Then, we randomly sampled 100 of them while establishing the initial taxonomy. Following ~\cite{xie2018people}, we assume each session to involve one topic. The query log analysis provided valuable data on real-world search behaviors, allowing us to identify prevalent search patterns and refine our understanding of user intents.} 
% ~(over 3,000 queries included)

\paragraph{Initial Taxonomy. }
The authors closely reviewed the survey responses and query logs to induce potential criteria for classifying user intents. Following previous work~\cite{rose2004understanding,xie2018people}, we took several iterations of the open-coded discussion before reaching an agreement. The initial version of the taxonomy is as shown in Figure~\ref{fig:taxonomy}~(the left one), which consists of six intents categorized by three criteria. The three criteria are the same as described in Section~\ref{sec:overview}. Specifically, the first criterion leads to three categories, i.e., Search for \textit{Particular Case(s)}, \textit{Learning}, and \textit{Analysis}, in this version. The \textit{Analysis} category denotes searching for writing some analytical reports, such as similar case search reports, which are sometimes required in the judicial process. At this point, we considered it as an independent search intent since it is a specialized task in judicial practice. The following section~(Section~\ref{subsec:verify}) will explain how we revise and verify the intent taxonomy.  

\subsection{Revisit and Verification}
% \subsection{Revising and Verifying}
\label{subsec:verify}
As illustrated in Figure~\ref{fig:procedure}, we conducted semi-structured interviews, collecting exhaustive feedback from experienced legal practitioners and making a qualitative analysis. Revisions were made accordingly. Following that, we conducted an editorial user study based on user survey responses and query logs, making \blue{a quantitative inspect}.

\subsubsection{Interview} 
\label{subsubsec:interview}
We conducted semi-structured interviews with four experienced legal workers separately, including one lawyer, one prosecutor, and two judges. \blue{They all work in Beijing. Three of them are men and one is a woman.} Although the face-to-face interview only involves a small sample, it could allow a more in-depth questioning and discussion, broadening and deepening the understanding on the research problem~\cite{thomas2019investigating}. The interviewees in our study were all well-experienced in legal practice and came from representative legal occupations. Each interview took about 30 minutes. Interviewees were compensated about 100 dollars for their participation. The audio was recorded for later analysis.

To begin with each interview, the interviewer introduced the proposed taxonomy in detail, including the three criteria, the hierarchical structure, and six intent categories, as shown in Figure~\ref{fig:taxonomy} and Table~\ref{tab:taxonomy}. Each interview was centered on two open questions plus a short series of follow-up questions.

The first question asked about the intent taxonomy's coverage and rationality. Here is an example\footnote{Interviews were all in Chinese. We show the translation in this paper. The exact wording varied for each interview.}. Firstly, we asked, \textit{What do you think of the coverage of this taxonomy? Can it cover all your information needs in legal search daily? If not, is there anything else that needs to be added?} Then, we followed with questioning about the concrete categories. For example, \textit{How about the XXX category? What do you think about the definition and characteristics of this category? } The follow-up question is a good trigger for open discussions. We collected rich comments and views on these intent categories from the perspectives of diverse legal occupations, which further helped us revise the taxonomy.  

The second question asked about the importance of different categories in the interviewee's daily search. For example, \textit{Among these intent categories, what do you think are more critical or occur more often in your daily search? And why?} We designed this question to obtain explicit feedback on the importance of different intents in the practice of legal case retrieval. Unlike user surveys or search logs, we could receive much more fine-grained explanations regarding this aspect despite the small data samples.

\paragraph{Results.}
After completing all the interviews, we analyzed the records. The main results are summarized as follows.
\begin{enumerate}
    \item Regarding the first question, all the interviewees stated that the proposed taxonomy has good coverage of daily needs in legal case retrieval. No more new categories were proposed. 
    \item Regarding the comments on each intent category~(i.e., the follow-up question), the \textit{Analysis} category attracted plenty of discussions. The lawyer and the judges, who usually dealt with such analytical reports~(e.g., similar case search reports), indicated that this type~(\textit{Analysis}) could be covered by the other categories mentioned earlier. Although it highly depends on the individual case, the underlying information need is still to learn about a specific legal problem~(e.g., \textit{Characterization} or \textit{Penalty} most of the time). The prosecutor interviewee indicated that he seldom had this type of intent. The potential situation he came up with is that when dealing with a difficult legal issue~(e.g., the \textit{Procedure}), he might also sum it up to an analytical report afterward, such as personal learning material. 
    \item Other comments on the concrete categories are centered on the categories under \textbf{Criterion3}. To be specific, the \textit{Characteriztion} category should also include the situations of innocence and those of non-prosecution (\textit{from the prosecutor}). Under the \textit{Procedure} intent, they usually search for the legal requirement related to jurisdiction or avoidance (\textit{from the lawyer}). Regarding the \textit{Penalty} intent, all the interviewees mentioned that it has attracted increasing attention in recent judicial practice, but meanwhile it is much harder to be satisfied in the current legal case search systems. Precision was especially emphasized under this intent. Last but not least, they all suggested that these three intents expect more diversified results than the \textit{Particular Case(s)} intent and meanwhile require higher precision and recall than the \textit{Interest} intent.
    \item Regarding the second question, all the interviewees suggested that the \textit{Characterization} and \textit{Penalty} are the most important and common in their daily search. Especially, the \textit{Penalty} was emphasized again by the prosecutor and the lawyer separately. Meanwhile, the prosecutor and the lawyer also mentioned that the \textit{Procedure} is highly significant. Although the \textit{Procedure} intent is less common than the above two categories in legal case search, it will be pretty valuable and, meanwhile, difficult if there is a need for case retrieval surrounding the procedure requirement. 
\end{enumerate}

\textit{Revisit and Revision.} Based on the interview responses, we had further iterative discussions on the taxonomy and finally reached an agreement on the revision as illustrated in Figure~\ref{fig:taxonomy}. To be specific, we removed the \textit{Analysis} category since it could be covered by the left intent categories. It would be better to view it as a context that triggers a legal case search rather than an independent intent search category. Furthermore, the first two authors re-coded the user survey and the sampled search logs that were used to establish the taxonomy in Section~\ref{subsec:establish}. As a result, the revised taxonomy still had good coverage. To summarize, we achieved a revised taxonomy composed of three criteria and five intent categories, as illustrated in Figure~\ref{fig:taxonomy}(b). Meanwhile, the in-depth discussions and exhaustive feedback help us further clarify the definitions of intent categories and also give us an qualitative view of the importance of different categories in practice. 

\subsubsection{Editorial User Study} 
\label{subsubsec:label}
To verify the revised taxonomy, we further conducted an editorial user study. In this study, we recruited three external legal experts to annotate the users' search intents in the user survey responses and query logs. The user survey responses and query logs are those described in Section~\ref{subsec:establish}. Unlike the establishing stage, we utilized all the sampled query logs (600 sessions in total) this time. The three annotators were all graduate students majoring in law and qualified in legal practice\footnote{They had passed the ``National Uniform Legal Profession Qualification Examination'' and had at least five years of law-related experience}. They all reported using legal case retrieval regularly and being familiar with current legal case search engines. They all signed a consent form before participating.

At the beginning of the study, we introduced the revised taxonomy in detail. We provided the annotators with the criteria, the taxonomy structure, the description, and examples of each intent category. In addition to the five intent categories, we provided another two choices for the annotators, \textit{Others}~(O) and \textit{Multi}~(M). The \textit{O} means that the search intent does not belong to any of the proposed categories. The \textit{M} denotes that the underlying intent seems to fall into multiple categories. For the additional two choices, we asked annotators to provide explanations for their choice. For example, the annotator needed to give what intent categories the search task might fall into if she selected \textit{M}.
The annotators were required to annotate the underlying search intent of each response in the user survey based on the answers to the three open questions and annotate the search intent of each session according to its queries. After all the annotators confirmed a good understanding of the taxonomy, they annotated the survey responses and query logs independently. 

It took about 1.5 hours and 7 hours on average for each annotator to annotate the user survey responses and query logs, respectively. Each annotator would be paid about \$12 for a one-hour annotation. As for label aggregation, we utilized the majority vote. In particular, if every annotator made different annotations for a sample, we tagged it as \textit{Multi}~(M).

\paragraph{Results.} The Fleiss's Kappa~\cite{fleiss1971measuring} $\kappa$ among three annotators is 0.62 in terms of the user survey annotation, reaching a substantial agreement~((0.61, 0.80)). As for the query log annotation, the $\kappa$ among three annotators is 0.58, reaching a moderate agreement~((0.41-0.60)). Compared with the survey where users described their search scenario explicitly, the query logs were vaguer for intent labeling and thus explained the slight drop in $\kappa$~\cite{su2018user,xie2018people}. Given the relatively high number of categories, the inner-annotator consistency is acceptable~\cite{xie2018people,bolotova2022non} for both datasets, suggesting that the taxonomy can be easily understood and distinguished.

\begin{table}
    \centering
    \caption{Distribution of search intent categories.}
    \label{tab:intent_dis}
    \begin{tabular}{lcc}
    \toprule
    \textbf{Intent Category} & \textbf{User Survey} & \textbf{Query Log} \\
    \midrule
    Particular Case(s) (\textit{PC}) & 7.27\% & 21.24\% \\
    Characterization (\textit{Ch}) & 50.00\% & 54.85\% \\
    Penalty (\textit{Pe}) & 10.91\% & 9.03\% \\
    Procedure (\textit{Pr}) & 6.36\% & 4.01\% \\
    Interest (\textit{In}) & 12.73\% & 0.17\% \\
    \midrule
    Others (\textit{O}) & 0.91\% & 0.67\% \\
    Multi (\textit{M}) & 11.82\% & 10.03\% \\
    \bottomrule
    \end{tabular}
\end{table}

Table~\ref{tab:intent_dis} shows the proportion of each intent category. As a result, less than 1\% search tasks were annotated as \textit{Others} in both user survey and query log datasets, indicating that the proposed taxonomy has good coverage of users' intents in legal case retrieval.

Regarding the five categories in the taxonomy, the general distributions are similar in both datasets, especially for the three intents classified by \textbf{Criterion 3}.
In particular, the \textit{Characterization} intent accounts for about 50\%, indicating it is a fundamental and common task in legal case retrieval. Consistently, recent research and benchmarks~\cite{shao2021investigating,ma2021lecard,shao2022understanding} designed tasks mainly based on this category. Meanwhile, the proportions of the \textit{Penalty} and the \textit{Procedure} are lower but still non-trivial compared with that of the \textit{Characterization}, which also aligned with the feedback collected in the interviews. These two are also primary tasks in the legal decision process. Besides, as the interviewees~(in Section~\ref{subsubsec:interview}) also pointed out, the need for \textit{Penalty} has been growing and \blue{is} increasingly important in recent years.  

Meanwhile, the distributions of \textit{Particular Case(s)} and \textit{Interest} vary in the two datasets. A higher proportion of \textit{PC} intent is observed in search logs while few \textit{Interest} intents are in the logs. We think that users' explicit responses in the survey can better reflect their real information needs, while the query logs are implicit indicators. Besides, the search engine itself might cause some bias in user preference. For example, if the search engine is not good at satisfying \textit{Interest} needs, users may not like to use it for this intent, and vice versa. According to the survey, we also noted that some users would use Web search engines rather than legal databases under the \textit{Interest} intent. 

\begin{figure*}[t]
\centering
\subfigure[User Survey]{
\includegraphics[width=.45\columnwidth]{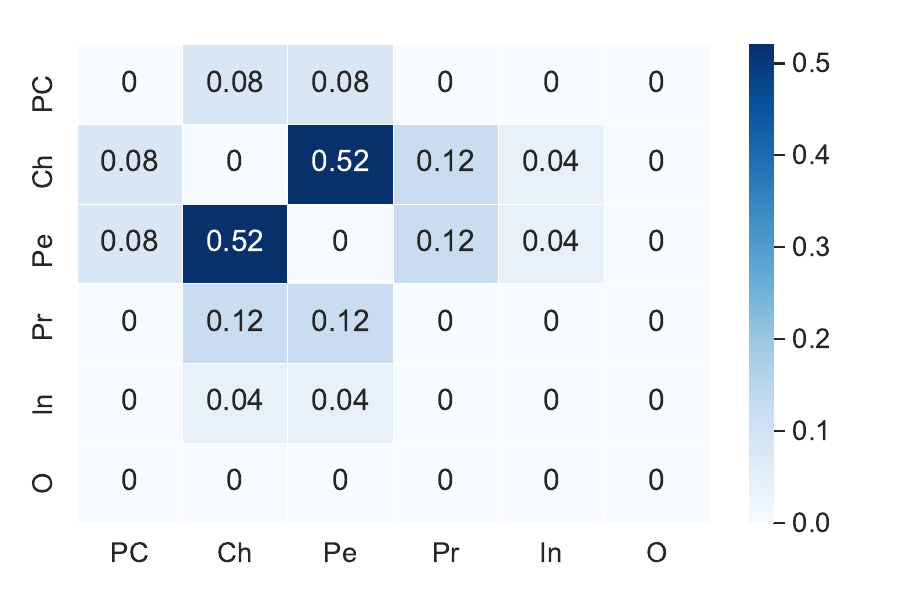}
\label{subfig:survey}
}
\subfigure[Query Logs]{
\includegraphics[width=.45\columnwidth]{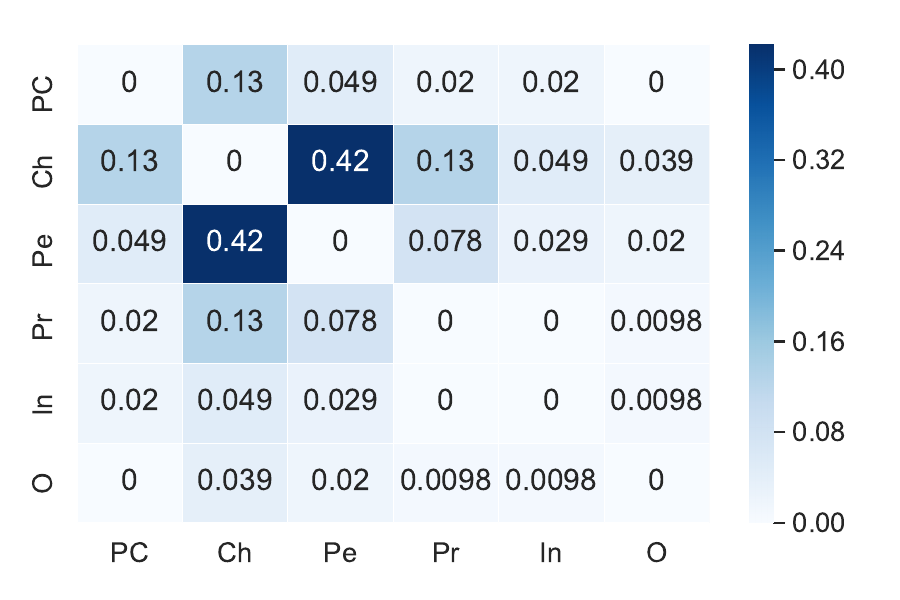}
\label{subfig:log}
}
\caption{Confusion matrix of the ``Multi'' in the user survey~(a) and query logs~(b). Each matrix is a symmetric one and the number in the grid denotes the normalized frequency of co-occurrence. } 
\label{fig:matrix}
\end{figure*}

\paragraph{Mixture Analysis.} Nearly 10\% of search tasks are tagged as \textit{Multi} in both datasets. Note that \textit{Multi} in Table~\ref{tab:taxonomy} consists of two parts, i.e., more than two annotators labeled as \textit{Multi} (7.27\% and 4.85\% in the survey and logs, respectively) or three annotators gave completely different labels (4.55\% and 5.18\% in the survey and logs, respectively). To deeply analyze it, we visualize the co-occurrence of different intents, as shown in Figure~\ref{fig:matrix}. For each sample belonging to the first part, we manually processed the annotators' explanations, from which we extracted all potential intents. We considered all annotated categories as possible intents for each sample belonging to the second part. Then, we count each pair in the possible intent set as one co-occurrence. For example, the intent set, ``Ch+Pr+Pe'', contributes once occurrence for the ``Ch-Pe'', ``Ch-Pr'', and ``Pe-Pr'' pairs, respectively. Numbers in Figure~\ref{fig:matrix} are normalized by the number of pairs.
% By reviewing these samples and annotators' explanations manually, we find that some tasks seem pretty ambiguous and might involve several possible intents simultaneously.

As shown in Figure~\ref{fig:matrix}, the pair of \textit{Characterization} and \textit{Penalty} is the one that co-occurs most frequently in both user survey and query log data, accounting for around 50\%, which suggests that users might search for both needs simultaneously. Meanwhile, we observe that the \textit{Procedure} usually co-occurs with the above two intents, which also aligns with the hierarchical structure of the intent taxonomy. Generally speaking, the query logs, where user intents could only be inferred implicitly, involve more types of co-occurrence of potential intents compared to the user survey. \blue{The results here suggest retrieval methods that explicitly recognize multi-intent queries are needed.}

\subsubsection{Summary} Based on the \textit{revisit and verification} stage composed of the interviews and editorial user study, we finalized the intent taxonomy, consisting of five categories, i.e., Search for \textit{Particular Case(s)}, \textit{Characterization}, \textit{Penalty}, \textit{Procedure}, and \textit{Interest}. We also provide quantitative insights into the search intent distributions in legal case retrieval, according to the intent annotations of user survey responses and query logs. 

\begin{table}[t]
\caption{\blue{Examples of User Study. The different questions reflect the corresponding type of intent. Users need to retrieve relevant cases regarding the question. }}
\begin{tabular}{p{0.65\columnwidth}p{0.35\columnwidth}}
\hline
\blue{Background} & \blue{Question}                             \\ \hline
\blue{\textbf{Criminal case:} The defendant Song was the son-in-law of the victim Li, and the two had a long-standing conflict. At 3:00 p.m. on January 20, 2019, the two clashed. Song rushed into the kitchen and casually picked up a knife and stabbed Li. After Li was stabbed 8 times, the knife suddenly broke. At the same time, Song's wife Chen just came home and went up to pull to stop Song. Seeing his wife Chen's grief, Song felt very regretful and sorry for his wife, so he held a knife to self-harm. Chen immediately called the police, Song knew that Chen was calling the police and did not stop. Because of the seriousness of the injury, the police immediately took Song to the hospital, the day began residential surveillance, by two police officers at the same time.} & \blue{\textbf{PC}:  The proceedings of Song.}

\blue{\textbf{Ch:} Is the defendant's act an intentional killing or intentional injury?}

\blue{\textbf{Pe:} Is the defendant's behavior a criminal suspension or an attempt to commit a crime?}

\blue{\textbf{Pr:} Can residential surveillance be treated as a prison term?} \\ \hline
\blue{\textbf{Civil case:} Zhang was running a beauty salon. between 2016 and 2019, Chen spent several times at Zhang's beauty salon and paid Zhang a total of RMB 7,700. Zhang informed Chen that all of the above items were of the best quality, but in reality they were all fake. The hyaluronic acid injected by Zhang for her caused Chen to develop hard lumps on her chin and chest. Chen filed a lawsuit in January 2020.}                                                   & \blue{\textbf{PC:} The proceedings of Zhang.}

\blue{\textbf{Ch:} Is Chen's claim based on product liability or consumer fraud liability?}

\blue{\textbf{Pe:} Is Chen's request for the return of \$7,700 and punitive damages supportable?}

\blue{\textbf{Pr:} Is the act committed by Zhang in 2017 time-barred?}      \\ \hline
\end{tabular}
\label{example}
\end{table}

\section{Search Behavior and Satisfaction}
\label{sec:behavior_satis}
To understand user search behavior and satisfaction under different search intents, we conducted a laboratory user study using the proposed taxonomy. In this section, we described the behavioral data collection process and addressed \textbf{RQ2} and \textbf{RQ3} with the collected data. 

\subsection{User Study}
\subsubsection{Tasks and Participants}
We designed three search tasks for each intent category. In each task, we provided a query case description as the background and a question specific to the intent category. \blue{Table ~\ref{example} shows a criminal example and a civil example.} The participant needed to retrieve relevant cases regarding questiones. All the tasks were adopted from the real cases to simulate the realistic search scenario. Similar to the previous study~\cite{shao2021investigating}, we anonymized the background case and removed the courts' opinions. \blue{So the query case contains only the basic facts.} All the tasks were designed to be of moderate difficulty to avoid impacts of task difficulty. \blue{Following ~\cite{shao2021investigating}, moderate difficulty cases are selected by experienced law professors.} The \textit{Interest} category was not included in the user study, since we could hardly simulate the search tasks triggered by an individual interest in a laboratory setting according to our prior study. In total, we designed 12 search tasks for the left four categories~(i.e., \textit{PC}, \textit{Ch}, \textit{Pr}, and \textit{Pe}) and an additional warm-up task.

We recruited 36 participants that were qualified\footnote{They passed the ``National Uniform Legal Profession Qualification Examination''} in legal practice. Specifically, eight were lawyers and the others were students in law school. They were native Chinese speakers and familiar with legal case retrieval. Considering the workload, we assigned each participant 3 main search tasks along with a warm-up training. On average, it took about 1.5 hours for each participant to complete the tasks. We carefully designed the assignment that each participant completed tasks of three different intents, and each task was completed by nine different participants. Tasks were shown in a random order to balance the order effects~\cite{lagun2014towards,shao2021investigating}.
A pivot study involving two additional participants was conducted ahead to ensure the experimental design worked well. 

\subsubsection{Procedure} First of all, we introduced the entire experimental procedure. After signing the consent form, the participant was directed to a warm-up task to get familiar with the experimental settings. Then, the participant moved on to the three main tasks. 
Each task consists of the following four steps.

\paragraph{Step1.} The participant was provided with the query case description and the question. She was instructed to search for relevant cases that could help her answer the question. After reading the case background and the question, the participant filled in a pre-search questionnaire to report her perceived task difficulty on a 5-point Likert-type scale. 

\paragraph{Step2.} The participant was directed to the experimental search engine. The participant could conduct legal case retrieval freely as she usually did, such as querying, clicking, turning pages, and so on. The participant could finish the search session once she found enough results or could not find more. 

\paragraph{Step3.} The participant was directed to a post-task questionnaire that contained two questions. The first is to ask for her perceived satisfaction regarding the entire search session on a 5-point scale. The second requires the participant to summarize the retrieved results and answer the task question. This question is to ensure the participant accomplished the search tasks seriously. 

\paragraph{Step4.} The participant was instructed to provide feedback for each query. Specifically, the issued queries, along with the questionnaires, would be shown successively. Regarding each query, the SERP and titles of clicked results~(if any) were also provided for reminding. The participant was instructed to report her satisfaction on a 5-point scale~(1:not at all, 5:satisfied) regarding this query and select the reasons for clicking on the results~(if any). The reasons for clicking were collected in the form of a multi-choice question. The options include \textit{relevance}, \textit{diversity}, \textit{authority}, \textit{timeliness}, \textit{region}, \textit{inspiration}, \textit{ranking}, and \textit{others}. The descriptions of options are as shown in Table~\ref{tab:option}. If the participant chose the \textit{others} option, she also needed to provide the potential factors that were not included in these options. 

\begin{table}
    \centering
    \caption{Descriptions of options in the user study~(Step4).}
    \label{tab:option}
    \begin{tabular}{p{0.15\columnwidth}p{0.7\columnwidth}}
    \toprule
    \textbf{Option} & \textbf{Description} \\
    \midrule
    Relevance & The relevance to the query. \blue{If the retrieved case satisfies the user's search intent, it is considered relevant otherwise irrelevant.} \\
    \midrule
    Diversity & The diversified content or opinions, e.g., providing different information or opinions beyond the existing one.\\
    \midrule
    Authority & The authority of the retrieved case, e.g., the court level involved in the case. \\
    \midrule
    Timeliness & The time-related factors of the retrieved case, e.g., the time that the case happened or was judged.\\
    \midrule
    Region & The region-related factors of the retrieved case, e.g., the region that the case happened or was judged.\\
    \midrule
    Inspiration & The inspiration of the result, e.g., providing ideas of identifying useful cases or formulating queries.\\
    \midrule
    Ranking & The ranking position of the result. \\
    \bottomrule
    \end{tabular}
\end{table}

\subsubsection{Experimental System} We developed an experimental platform using Django where the participants completed the entire study procedure. As for the experimental search engine, \blue{we redirected to a commercial legal case retrieval engine~\footnote{\url{https://ydzk.chineselaw.com/case}}}. Query suggestions and advertisements were filtered. It had been confirmed in advance that the search system would not do personalization. We developed a customized chrome extension to log user behavior and examined pages. 

\subsubsection{Dataset} We collected 108 valid search sessions~(843 queries included) for the 12 tasks from 36 participants after the quality check. The dataset contained rich behavioral data~(e.g., queries, clicks, hovers, and timestamps) and users' explicit feedback~(e.g., satisfaction and click-through reasons). 
The dataset will be open to the public after acceptance. 

\subsection{Search Behavior under Different Intents}
Regarding \textbf{RQ2}, we investigated users' search behavior under different intents based on the collected behavioral data. 

The search tasks were designed to be of similar difficulty across different intents to avoid the potential influences of task difficulty~\cite{shao2021investigating}. This design is also verified by users' feedback on task difficulty in \textit{Step1}. The average task difficulty is 2.5, and no significant differences were observed across intents~($p>0.6$). 

Search intents are considered as independent variables. We follow the hierarchical structure in Figure~\ref{fig:taxonomy} for an in-depth investigation. Specifically, we first group sessions into the \textit{Particular Case(s)} and \textit{Learning} categories according to \textbf{Criterion 1}. Then, we apply \textbf{Criterion 3} to sessions in \textit{Learning} and group them into the \textit{Characterization}, \textit{Penalty}, and \textit{Procedure} categories. We investigate user behavior in legal case retrieval from multiple aspects, including task events, click, hover, and dwell time. Non-parametric statistical tests~(Kruskal-Wallis test~\cite{kruskal1952use}) are utilized since these measures have non-normal distributions~(K-S test). The p-values are calibrated through Bonferroni-Holm adjustment~\cite{holm1979simple} within the behavioral group to counteract the multiple comparison problem~\cite{fuhr2018some}. Results are given in Table~\ref{tab:behavior}. 

\begin{table*}[t]\small
    \centering
    \caption{Differences in user behavior with different search intents. PC/Le/Ch/Pe/Pr denotes for Particular Case(s), Learning, Characterization, Penalty, and Procedure, respectively. ``*/**/***'' indicates a significant difference at $p<0.05/0.01/0.001$ level (after Bonferroni-Holm correction). }
    \label{tab:behavior}
    \begin{tabular}{p{0.17\columnwidth}p{0.25\columnwidth}ccccccc}
    \toprule
    \multirow{2}{*}{\textbf{Group}} & \multirow{2}{*}{\textbf{Behavioral Measure}} & \multicolumn{3}{c}{\textbf{Criterion 1}} & \multicolumn{4}{c}{\textbf{Criterion 3}} \\
    \cmidrule(lr){3-5} \cmidrule(lr){6-9}
    & & \textbf{PC} & \textbf{Le} & sig. & \textbf{Ch} & \textbf{Pe} & \textbf{Pr} & sig. \\
    \midrule
    \multirow{3}{*}{\textbf{Task Events}} & \# query per session & 6.346 & 6.870 & -- & 4.615 & \textbf{9.000} & 7.000	& * \\ 
    %  & \textbf{\% specification} & \textbf{0.5590} & 0.4287 & ** & 0.4207 & 0.4480 & 0.4163 & -- \\
    %  & \% generalization & 0.1647 & 0.1697 & -- & 0.1388 & 0.1703 & 0.1953 & -- \\
    %  & \textbf{\% substitution } & 0.2763 & \textbf{0.4015} & * & 0.4405 & 0.3817 & 0.3884 & -- \\
    % \hline
    % \multirow{2}{*}{\textbf{Pages}} 
     & \# pages per session & 11.17 & 13.78 & -- & 10.24 & \textbf{17.74} & 13.32 & ** \\
     & \# search depth in pages & \textbf{1.309} & 1.063 & ** & 1.087 & 1.052 & 1.029 & -- \\
    \midrule
    \multirow{4}{*}{\textbf{Click}} & \# clicks per session & 5.615 & 5.886 & -- & 3.435 & \textbf{7.593} & 4.565 & *** \\
    & min click rank per query & 1.269 & \textbf{1.963} & *** & 1.630 & 1.913 & 2.500 & -- \\
    & avg click rank per query & 2.823 & \textbf{3.551} & * & 3.264 & 3.288 & 
\textbf{4.189} & * \\
    & \% sats click per query & 0.3294 & \textbf{0.6485} & *** & \textbf{0.7150} & 0.7090 & 0.5214 & ** \\
    % & \textbf{avg skipped results per query} & 0.7613 & \textbf{1.853} & ** & 2.259 & 1.660 & 1.742 & -- \\
    \midrule
    \multirow{5}{*}{\textbf{Hover}} & \# hovers per session & 50.78 & 46.73 & -- & 33.61 & \textbf{59.44} & 46.95 & * \\
    & min hover rank per query & 1.142 & 1.075 & -- & 1.095 & 1.051 & 1.088 & -- \\
    & avg hover rank per query & 3.059 & 3.170 & -- & 3.356 & 3.171 & 3.062 & -- \\
    & avg hover time (seconds) per query & 2.221 & \textbf{2.790} & ** & 2.722 & 2.938 & 2.684 & -- \\
    & P(click|hover) per query & 0.2044 & 0.1817 & -- & \textbf{0.2099} & 0.1927 & 0.1484 & ** \\
    \midrule
    \multirow{3}{*}{\textbf{Dwell Time}} & task time (seconds) per session & 379.5 & \textbf{551.3} & ** & 438.3 & \textbf{702.7} & 425.8 & ** \\
    & \% SERP time per session & \textbf{0.6042} & 0.4218 & *** & 0.3868 & 0.4255 & 0.4637 & -- \\
    & avg click dwell (seconds) per query & 28.76 & \textbf{53.37} & *** & \textbf{66.83} & 53.10 & 43.41 & ** \\
    \bottomrule
    \end{tabular}
\end{table*}

\paragraph{Task Events.} Comparing the \textit{Particular Case(s)} and \textit{Learning} intents, the general numbers of issued queries and visited pages are similar. \blue{We suppose this may be due to the fact that users tend to prefer simple keyword expressions when searching for particular cases. When the satisfactory case cannot be found using simple keywords, the user will further enrich the query description.} Regarding search depth in pages~(the number of SERP pages a user browses per query~\cite{su2018user}), users turn pages more often under the \textit{Particular Case(s)} intent. \blue{
On one hand, this reflects the fact that current user search habits do not facilitate the rapid identification of specific cases. On the other hand, the difference reflects the requirement for both high precision and high recall given the \textit{Particular Case(s)} intent.} Meanwhile, we observe significant differences in the number of queries and pages when comparing among \textit{Characterization}, \textit{Penalty}, and \textit{Procedure}. More queries and pages are examined under the \textit{Penalty} intent, indicating a higher search effort. The results could be interpreted by the legal characteristics of the \textit{Penalty} category that the underlying information need is usually more specific and precise.     

\paragraph{Click.} We observe significant differences in all the query-level click-through measures between the \textit{Particular Case(s)} and \textit{Learning}. The difference in search purpose~(\textbf{Criterion 1}) might lead to remarkably different examination patterns within a query. The differences in \textit{min} and \textit{avg} clicked positions indicate that users with \textit{Learning} intent seem more patient and careful with the returned results. 
\blue{When using a 30-second threshold to determine a satisfactory click, a higher proportion of clicks under the \textit{Learning} intent category meet this satisfaction criteria.} Comparing among the intents categorized by \textbf{Criterion 3}, click-through behavior measures also vary significantly. The \textit{Penalty} involves the most clicks within a session, which is consistent with the analysis of the above task event measures. 
Furthermore, according the \textit{\%sats click}, users seem to be least satisfied with the results under the \textit{Procedure} intent. Different from \textit{Characterization} and \textit{Penalty}, the requirement under \textit{Pr} is based on the procedural law and the corresponding relevance criteria might also differ. Without understanding the user intent, the existing retrieval systems might hardly resolve this kind of information need well. 
 % at session-level and query-level

\paragraph{Hover.} Following previous works~\cite{chen2017meta,shao2021investigating}, we utilize hover measures to reflect users' examination of the results shown on SERPs~(e.g., snippets). They could capture more behavioral information than click-through measures but might involve more noise. Specifically, we view hover-through as a signal of a preliminary examination, which involves less examination effort than click-through. Comparing \textit{Particular Case(s)} and \textit{Learning}, the main difference lies in the average hover time on results. In the context of \textit{Learning} intent, it would take users more time to examine and understand the result content. Among the three categories under \textit{Learning}, \blue{\textit{Penalty} involves the most hovers, which may indicate a higher search effort needed under this intent.} Meanwhile, \textit{P(click|hover)}~(the probability of a result to be clicked given hovered~\cite{shao2021investigating}) is significantly lower in the \textit{Procedure} intent than the others, indicating the user might skip more irrelevant results based on her preliminary judgments. It also suggests that the existing result list might not well satisfy the information need of \textit{Procedure}. 

\paragraph{Dwell Time.} Although \textit{Particular Case(s)} and \textit{Learning} tasks involve similar numbers of queries and visited pages, the total task time is significantly longer in \textit{Learning}. \blue{In particular, users spent more time on SERPs under the \textit{Particular Case(s)} intent, while they spend more time on clicked case documents under the \textit{Learning} intent.} Compared among \textit{Characterization}, \textit{Penalty}, and \textit{Procedure}, unsurprisingly, the \textit{Penalty} tasks tend to take much more task time. \blue{Specifically, it took remarkably more time to examine the clicked results under the \textit{Characterization} intent}. Since the target information regarding the \textit{Characterization} intent is usually broader in scope, users may need to read more contents in a case document to understand the entire case well. 

\paragraph{Summary.} Users' search behavior in legal case retrieval varies significantly with search intents regarding various aspects. Comparing \textit{Particular Case(s)} and \textit{Learning} intents, users seem more patient and spend more time examining the content of case documents in \textit{Learning}. With the requirement for high precision and recall, users with \textit{Particular Case(s)} intent might put more effort into exploring the SERPs. Among \textit{Characterization}, \textit{Penalty}, and \textit{Procedure}, the \textit{Penalty} tasks always involve the most search effort. Meanwhile, we observe that users with \textit{Procedure} intent seem quite patient but less satisfied with the system results. 

% Meanwhile, users tend to spend more time reading the clicked case document under the \textit{Characterization} tasks.

% \vspace{-2mm}
\subsection{User Satisfaction under Different Intents}
User satisfaction is a key concept in information retrieval systems, measuring the fulfillment of a user's information need~\cite{su1992evaluation,kelly2009methods}.
To answer \textbf{RQ3}, we first investigate how user satisfaction distributes under each intent and the influential factors according to users' explicit feedback. Furthermore, towards the evaluation of legal case retrieval, we attempt to measure user satisfaction with online metrics based on implicit signals. 
% in different search intent scenarios.

\subsubsection{Explicit Feedback}
We observe significant differences in user satisfaction feedback across search intents. The average query satisfaction under each intent~(i.e., \textit{PC}, \textit{Ch}, \textit{Pe}, \textit{Pr}) is 3.441, 3.207, 3.127, and 2.950, respectively~($p<0.01$). 
Specifically, users perceive significantly higher satisfaction in the \textit{Particular Case(s)} scenario than in the \textit{Learning}~($p<0.001$). Comparing the three categories within the \textit{Learning}, the difference is mainly between \textit{Procedure} and the others. Users seem not well satisfied in the \textit{Procedure} context. In that case, the legal case retrieval systems need to put more effort into satisfying users' \textit{Learning} tasks, and especially, attach due importance to the \textit{Procedure} ones.

\begin{figure}
    \centering
    \includegraphics[width=0.7\columnwidth]{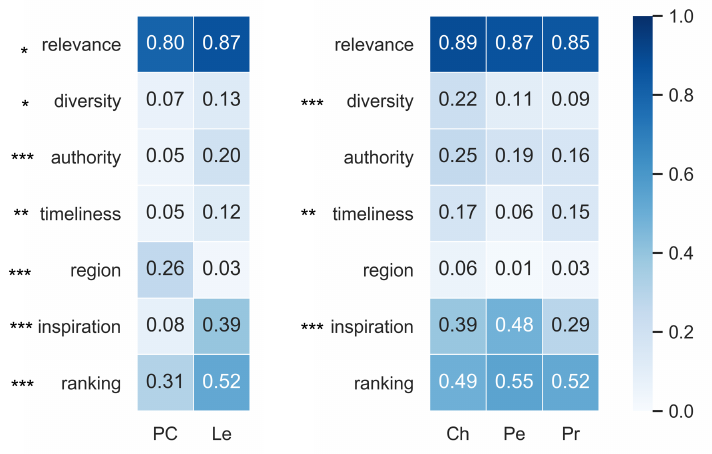}
    \caption{The distribution of click reasons across different intents. The number in the grid denotes the proportion of the users that select the factor under the intent, correspondingly. ``*/**/***'' indicates the statistical significance at $p<0.05/0.01/0.001$ level by one-way ANOVA, respectively. }
    \label{fig:click_reasons}
\end{figure}

Further, we inspect the potential \blue{factors influencing} user satisfaction under different search intents based on users' feedback on reasons for click-through~(collected in \textit{Step4)}. Following previous studies~\cite{dan2016measuring, wu2019influence}, we consider the click as an important indicator of satisfaction. 
Firstly, no new factors outside of our seven options were proposed by the participants in our study. Therefore, we focus on the seven factors that we provided in the user study, i.e., \textit{relevance}, \textit{diversity}, \textit{authority}, \textit{timeliness}, \textit{region}, \textit{inspiration}, \textit{ranking}. Figure~\ref{fig:click_reasons} shows the distributions of these factors. 

Across all search intents, \textit{relevance} is always the main concern. However, beyond relevance, users pay attention to different aspects under different search intents. 
We observe that users may emphasize different aspects of a legal case document when search under the intents of Particular cases and Learning. \blue{Locality is more important when searching for particular cases, than we searching to satisfy an information need. On the contrary, when users search for \textit{Learning}, they tend to care about other properties of the case contents, such as \textit{authority}, \textit{diversity}, and \textit{timeliness}.} It is worth mentioning that \textit{inspiration}, which means the result could inspire users to formulate better queries or find other cases, is emphasized under the \textit{Learning} intent more often. We believe that \blue{the inspiration} could help the user's exploratory search process. Moreover, the system ranking is more critical in the \textit{Learning} tasks. Considering the larger result set and higher effort in examining results, we think users would rely more on the system rankings to identify better results. It also highlights the importance of optimizing top-ranked results in legal case retrieval, especially for the \textit{Learning} tasks, although users might be more patient than in general Web search~\cite{shao2021investigating}. 
% the potential relevant results might be broader. Therefore,

Significant differences mainly lie in \textit{diversity}, \textit{timeliness}, and \textit{inspiration} when we compare among \textit{Characterization}, \textit{Penalty}, and \textit{Procedure}. In particular, the \textit{Characterization} intent requires a much higher level of result \textit{diversity} than the others. The \textit{inspiration} factor is more influential in the search intents regarding the substantive law, especially in the \textit{Penalty} intent. Since users tend to put the most search effort into the \textit{Penalty} tasks, the results with high inspiration would benefit the user's exploratory process. Meanwhile, results for the issues under the procedural law are usually within a more definite scope than those under the substantive law, \blue{which may be why users care} less about result \textit{diversity} and \textit{inspiration} under the \textit{Procedure} intent. 

To sum up, users pay attention to different factors beyond relevance~(e.g, \textit{diversity}, \textit{region}, \textit{inspiration}, etc.) given different search intents. The results also shed light on the optimization directions for legal case search systems to promote user satisfaction under different search intents. 

\subsubsection{Implicit Signals}
\begin{table}\small
    \centering
    \caption{Online Metrics and their descriptions. }
    \label{tab:online_metrics}
    \begin{tabular}{p{0.1\columnwidth}p{0.25\columnwidth}p{0.55\columnwidth}}
    \toprule
    \textbf{Group} & \textbf{Metrics} & \textbf{Description} \\
    \midrule
    \multirow{3}{*}{Click} & UCTR & a binary variable indicating whether there was a click or not \\
      & QCTR & the number of clicks \\
      & MaxRR/MinRR/MeanRR & maximum/minimum/mean reciprocal ranks (RR) respectively \\
    \midrule
    \multirow{5}{*}{Dwell} & SumClickDwell & the sum of click dwell time \\
     & AvgClickDwell & the average of click dwell time \\
     & QueryDwell & dwell time of the query session\\
     & TimeToFirstClick & time delta between the start of the query and the first click \\
     & TimeToLastClick & time delta between the start of the query and the last click \\
    \bottomrule
    \end{tabular}
\end{table}

Evaluation plays an essential role in IR research, which measures how well the search system satisfies users' information needs. In contrast to offline evaluation metrics that rely on external relevance judgments, online metrics calculated based on behavioral logs~(implicit signals) are cheaper and widely adopted in current search engines for system evaluation. Although the evaluation metrics designed for legal case search are still under investigation, this paper focuses on the performance of some common metrics generally applied to diverse search scenarios, taking user satisfaction as the ``golden standard''~\cite{al2007relationship,chen2017meta,zhang2018well}. To be specific, we conduct a correlation analysis to investigate how existing online metrics could measure user satisfaction, especially under different legal search intents. 
Following previous research~\cite{zhang2018well, chen2017meta}, we inspect the popular click-based and dwell-based metrics. Table~\ref{tab:online_metrics} shows the online metrics that we use in this paper and their definitions. The Pearson's correlation coefficients between these online metrics and user satisfaction are shown in Table~\ref{tab:online_metrics_corr}. 

% Specifically, \textit{UCTR} and \textit{QCTR} denote whether the user clicks any result and the total number of clicks respectively. \textit{MaxRR}/\textit{MinRR}/\textit{MeanRR} denotes the maximum/minimum/average value of the reciprocal ranks of clicks. In the dwell-based metrics, \textit{TimeToFirstClick}/\textit{TimeToLastClick} denotes time interval between the start of search and the first/last click. 

% Online metrics

\begin{table}
    \centering
    \caption{Pearson's correlation between online metric and user satisfaction under different intents. * indicates the correlation is significant at $p<0.001$. }
    \label{tab:online_metrics_corr}
    \begin{tabular}{llcccc}
    \toprule
    \textbf{Group} & \textbf{Metric} & \textbf{PC} & \textbf{Ch} & \textbf{Pe} & \textbf{Pr} \\
    \midrule
    \multirow{5}{*}{\textbf{Click}} & UCTR & 0.2815* & \textbf{0.4608}* & 0.2238* & 0.3927* \\
     & QCTR & 0.2567* & \textbf{0.4447}* & 0.2943* & 0.3581* \\
     & MaxRR & 0.2532* & \textbf{0.4324}* & 0.2300* & 0.3681* \\
     & MinRR & 0.1475 & \textbf{0.2990}* & 0.0818 & 0.2504* \\
     & MeanRR & 0.2040 &	\textbf{0.3820}* & 0.1572 & 0.3234* \\
    \midrule
    \multirow{5}{*}{\textbf{Dwell}} & SumClickDwell & 0.3162* & \textbf{0.4748}* & 0.3022* & 0.3351* \\
     & AvgClickDwell & 0.2821* & \textbf{0.4362}* & 0.2539* & 0.3155* \\
     & QueryDwell & 0.1903 & 0.2067 & 0.1679 & \textbf{0.2875}* \\
     & TimeToFirstClick & 0.0056 & -0.0152 & -0.2045 & -0.0520 \\
     & TimeToLastClick & 0.3312 & 0.1819 & \textbf{0.3419}* & 0.1670 \\
    \bottomrule
    \end{tabular}
\end{table}
% Diverse offline and online evaluation metrics have been proposed and applied in Web search. 

\paragraph{Click-base Metrics.} \textit{UCTR} and \textit{QCTR} correlate significantly and positively with user satisfaction across all intent categories. Unlike the negative correlations in web search~\cite{chen2017meta}, users usually need to examine the case document to satisfy information needs, and thus more interactions with results are desired. Comparing among the search intents, the correlations with user satisfaction become weaker when the user's information need is relatively more specific~(e.g., \textit{Particular Case(s)} and \textit{Penalty}). Similar trends can also be observed in metrics based on click-through ranks. Specifically, \textit{MinRR} and \textit{MeanRR} can not well measure user satisfaction in the \textit{Particular Case(s)} and \textit{Penalty} scenarios. Only \textit{MaxRR}, indicating the top rank of click, has significant correlations with user satisfaction under all search intents.

\paragraph{Dwell-based Metrics.} \textit{SumClickDwell} and \textit{AvgClickDwell} have significant correlations with user satisfaction under all search intents. More time spent on examining case documents is a positive signal in legal case retrieval. However, \textit{QueryDwell}, calculated based on the query's total dwell time, only correlates significantly with user satisfaction given the \textit{Procedure} intent. Meanwhile, \textit{TimeToLastClick} seem more suitable to measure user satisfaction under the intents that involve relatively specific information needs, such as \textit{Penalty} and \textit{Particular Case(s)}~($p=0.009$). 

\blue{In summary, online metrics demonstrate varying performances when used as indicators of user satisfaction. Given the diversity of search intents, it is essential to reconsider the extent to which a metric can accurately reflect user satisfaction and effectively evaluate the system.}

% \subsection{Summary}
% In summary, there are remarkable differences in users' perception of satisfaction under different search intents, in terms of satisfaction distribution and influential factors. It is more difficult to satisfy users' \textit{learning} intent, especially the \textit{Pr}. Compare with \textit{PC}, users pay more attention to result diversity, authority, timeless, inspiration, and rankings under \textit{Le}. Diversity and inspiration are relatively desirable regarding the \textit{Ch} and \textit{Pe} tasks, respectively. On the other hand, online metrics achieve distinct performances in terms of measuring user satisfaction. 

\section{Applications}
\label{sec:application}
In this section, we attempt to apply the intent taxonomy to two critical downstream IR tasks to answer \textbf{RQ4}~(\textit{How can the taxonomy benefit downstream tasks in legal case retrieval}~), including satisfaction prediction and result ranking. 

\subsection{Satisfaction Prediction}
We attempt to predict user satisfaction with behavioral signals. In particular, we investigate the application of the intent taxonomy to this task from multiple perspectives. First, we inspect the performance of different behavioral signals in satisfaction prediction under different search intents. Second, we build an intent-aware model for satisfaction prediction. 

\begin{table}\small
    \centering
    \caption{Behavioral features used in satisfaction prediction.} 
    \label{tab:features}
    \begin{tabular}{p{0.2\columnwidth}p{0.5\columnwidth}c}
    \toprule
    \textbf{Feature Group} & \textbf{Feature Description} & \textbf{Numbers} \\
    \midrule
    \multirow{3}{*}{Click} & the number of clicks; & \multirow{3}{*}{5} \\
     & the click-through rate; & \\
     & maximum/minimum/mean reciprocal ranks of clicks; & \\
    \midrule
    \multirow{4}{*}{Hover} & the number of hovers; & \multirow{4}{*}{6} \\
     & the probability of being clicked given hovered; & \\
     & average of skipped results between hovers; & \\
     & maximum/minimum/mean ranks of hovers; \\
    \midrule
    \multirow{4}{*}{Dwell} & dwell time on SERP/Landing Pages; & \multirow{4}{*}{5} \\
     & time to first click; & \\
     & average of dwell time on hovered results; & \\
     & average of dwell time on clicked results; & \\
    \midrule
    \multirow{4}{*}{Query} & the length of query (in characters); & \multirow{4}{*}{4} \\
     & the number of query terms; & \\
     & the ratio of unique terms; & \\
     & the number of visited pages; & \\
    \bottomrule
    \end{tabular}
\end{table}

\subsubsection{Features}
User behavior has been popularly utilized to predict satisfaction in varied search scenarios, such as Web search~\cite{kim2014comparing}, product search~\cite{su2018user}, and image search~\cite{wu2019influence}. \blue{However, there is limited research dedicated to constructing models for predicting user satisfaction in legal case retrieval.} Referring to previous works~\cite{kim2014comparing,su2018user,wu2019influence} and preliminary analyses in the former sections, we extracted four groups of behavioral features~(20 in total), as shown in Table~\ref{tab:features}. Features in the \textit{Click}, \textit{Hover}, and \textit{Dwell} groups are the same as described in Section~\ref{sec:behavior_satis}. In the \textit{Query} group, we mainly utilized features that potentially reflect the overall complexity of this search through text statistics and browse pages. Note that we only used implicit signals~(logged behavior) in this task and did not include any explicit feedback, considering that explicit feedback is rather expensive to collect in practice. 

\subsubsection{Experimental Settings}
The behavioral dataset we used was collected in the user study as described in Section~\ref{sec:behavior_satis}. Following previous research~\cite{su1992evaluation,wu2019influence}, we mapped the 5-level satisfaction scale to a binary indicator (dissatisfied: 1\&2\&3, satisfied: 4\&5) and treated satisfaction prediction as a binary classification task. Prediction performance was evaluated by AUC considering the imbalanced distribution of labels~(dissatisfied: 470, satisfied: 345). We applied a gradient boosting decision tree model implemented by CatBoost~\cite{prokhorenkova2018catboost}, which can support both numerical and categorical features simultaneously and achieve great quality stably without parameter tuning. We considered two types of experimental settings, i.e., satisfaction prediction on the tasks of each intent and of all intents. Specifically, in the latter setting~(denoted as ``All Tasks'' in Table~\ref{tab:satisfaction prediction}, we compared the performance of intent-agnostic and intent-aware models. The intent-agnostic models are built based on the behavioral features listed in Table~\ref{tab:features} and trained on the tasks of all intents. The intent-aware models added the intent category to the behavioral features and trained on the same data of the intent-agnostic models. 
Experiments were all conducted on 5-fold cross-validation. 

\begin{table}
    \centering
    \caption{Satisfaction prediction performance measured by AUC. Results in boldface denote the best feature group and the best performance for each column. }
    \label{tab:satisfaction prediction}
    \begin{tabular}{lcccccc}
    \toprule
     & \multicolumn{4}{c}{\textbf{Tasks per Intent}} & \multicolumn{2}{c}{\textbf{All Tasks}} \\
      \cmidrule(lr){2-5} \cmidrule(lr){6-7}
     & PC & Ch & Pe & Pr &  \textit{Intent-agnostic} & \textit{Intent-aware} \\
    \midrule
    Click & 0.6314 & 0.7135 & 0.6025 & 0.5901 & 0.6150 & 0.6366 \\
    Hover & 0.6216 & 0.7145 & 0.6221 & 0.6261 & 0.6536 & 0.6523 \\
    Dwell & 0.5893 & \textbf{0.7255} & \textbf{0.6776} & 0.6395 & \textbf{0.6766} & \textbf{0.6918} \\
    Query & \textbf{0.6409} & 0.6831 & 0.5898 & \textbf{0.6685} & 0.5854 & 0.6187 \\
    \midrule
    All Features & \textbf{0.6996} & \textbf{0.7557} & 0.6648 & 0.6294 & 0.6728 & \textbf{0.7020} \\
    \bottomrule
    \end{tabular}
\end{table}

\subsubsection{Prediction Results}
Results are as shown in Table~\ref{tab:satisfaction prediction}. According to the prediction performance on ``tasks per intent'',  we observe differences in the performance of behavioral features under different search intents. Specifically, the \textit{Dwell} features achieve the best performance under the \textit{Characterization} and \textit{Penalty} tasks, while the \textit{Query} features are more effective under the \textit{Particular Case(s)} and \textit{Procedure} intents. Furthermore, combining all kinds of features does not always lead to improvements. For instance, the combination of all features achieves the best performance only under the \textit{Particular Case(s)} and \textit{Characterization} intents. However, the combination may lead to a drop under the other intents, especially in the \textit{Procedure} tasks. Meanwhile, comparing the prediction performance under different intents, user satisfaction under the \textit{Procedure} intent seems the most difficult to model, which might need further effort to optimize. The results suggest that different types of behavioral signals should be utilized for satisfaction prediction when the search intent varies. 

According to the performance comparison on ``all tasks'', the intent-aware model performs better than the intent-agnostic model most of the time, given different features. In particular, when using the \textit{Dwell} features and combination of features (\textit{All Features}), which perform relatively better than other feature groups, 
\blue{the intent-aware methods consistently demonstrate significant improvements~(t-test, $p<0.05$) in performance.} Specifically, the combination of behavioral features achieves the best performance with the search intent provided. The results suggest that involving search intent categories can contribute to improvements in satisfaction prediction performance.

\subsection{Ranking}
Ranking is a core task for IR. In this section, we exploited the widely adopted Learning to Rank~(LTR)~\cite{liu2009learning,li2023better,chen2023thuir} and attempted to integrate the proposed intent taxonomy into this task. 

\subsubsection{Model} We follow the intent-aware ranking adaption framework~\cite{glater2017intent} to integrate search intents with ranking models. To be specific, the probability that a result $r$ satisfies the query $q$ is calculated as, 
\begin{equation}
\label{equ:intent-aware}
    P(r|q) = \sum_{i \in I} P(i|q)P(r|q,i)
\end{equation}
, where $I$ denotes the intent set. $P(r|q, i)$ is denotes the probability that the result $r$ satisfies the query $q$ under the intent $i$ and is calculated by a intent-specific ranking module. Similar to~\cite{glater2017intent}, the intent-specific ranking module is an LTR model optimized for a specific intent. $P(i|q)$ denotes the probability of intent $i$ given the query $q$. In our study, $P(i|q)$ works as an indicator~($P(i|q) \in \{0, 1\}$) indicates which intent the query belongs to. We acknowledge that this simplification would be a limitation since it dismissed the mixture of multiple intents. However, developing complicated intent-aware ranking models is beyond this paper's main concern, and we leave it for future work.

\subsubsection{Experimental Settings} We utilized the sampled query logs with intent annotations as described in Section~\ref{subsec:establish}. We filtered out the data that were aggregated as \textit{Multi} to avoid noise and the data in the \textit{Others} or \textit{Interest} categories due to the data sparsity. The clicked results were viewed as relevant, and the left were regarded as irrelevant. We filtered out the queries without any clicks. After filtering, we ended up with 525 search sessions under four intents, consisting of 1,177 queries. Case documents were downloaded. 
We extracted content-based features that have been commonly used in the LTR literature~\cite{liu2009learning}, including average term frequency~(TF), average inverse document frequency~(IDF), average TF-IDF, BM25 score, and cosine similarity based on TF-IDF vectors. All the models in the experiment used the same feature set. \blue{Regarding the learning algorithm, we employed three ranking algorithms: LambdaMART~\cite{wu2008ranking}, RankBoost~\cite{freund2003efficient}, and AdaRank~\cite{xu2007adarank}. These algorithms cover point-wise, pair-wise, and list-wise ranking methodologies.} For each ranking algorithm, we compared the performance between the intent-agnostic model and the intent-aware one. To be specific, the intent-agnostic model was trained on the queries of all intent categories. Under the intent-aware framework, the intent-specific module was trained based on the queries under the corresponding intent. The final ranking score was calculated according to formula~(\ref{equ:intent-aware}). Both intent-agnostic and intent-aware models were tested on the same testing set, a mixture of queries under various intents. The algorithms were implemented by RankLib\footnote{\url{https://sourceforge.net/p/lemur/wiki/RankLib/}}. Parameters were set as default. We performed a five-fold cross-validation. In each cross-validation round, we used 10\% of the train data as the validation set and optimized NDCG@10. The final performance was evaluated by NDCG@5, NDCG@10, NDCG@15, and MAP. Table~\ref{tab:ranking} shows the average performance. 

\begin{table}
    \centering
    \caption{Comparison of intent-agnostic and intent-aware ranking models. Results in boldface denote the winner performance given each LTR model.}
    \label{tab:ranking}
    \begin{tabular}{lcccc}
    \toprule
     & \textbf{NDCG@5} & \textbf{NDCG@10} & \textbf{NDCG@15} & \textbf{MAP} \\
    \midrule
    AdaRank & 0.3811 & 0.4757 & 0.5361 & 0.3951 \\
    \quad w/ intent-aware & \textbf{0.4444} & \textbf{0.5270} & \textbf{0.5801} & \textbf{0.4471} \\
    \quad improv. & +16.6\% & +10.9\% & +8.21\% & +13.2\% \\
    \midrule
    LambdaMART & 0.4936 & 0.5753 & 0.6105 & 0.4852 \\
    \quad w/ intent-aware & \textbf{0.5329} & \textbf{0.6033} & \textbf{0.6354} & \textbf{0.5208} \\
    \quad improv. & +7.96\% & +4.87\% & +4.08\% & +7.34\% \\
    \midrule
    RankBoost & 0.5711 & 0.6446 & 0.6704 & 0.5648 \\
    \quad w/ intent-aware & \textbf{0.5820} & \textbf{0.6524} & \textbf{0.6777} & \textbf{0.5737} \\
    \quad improv. & +1.91\% & +1.21\% & +1.09\% & +1.58\% \\
    \bottomrule
    \end{tabular}
\end{table}

\subsubsection{Results} As shown in Table~\ref{tab:ranking}, integrating search intents into ranking improves the performance of existing intent-agnostic ranking algorithms in legal case retrieval. \blue{The intent-aware models trained ranking models separately based on the training data with different intents. If user's needs and concepts of relevance do not vary across different intents in the developed intent taxonomy, the consideration of intents in the intent-aware models would only add noise to the training data, and make the final ranking models perform worse because the split of training data would make each model has less data for parameter optimization. However, as shown in Table 11, even with less training data for each intent ranking model, the intent-aware models still outperform the intent-agnostic models that trained with the full data. This demonstrates that users who submitted the queries with different intents under our intent taxonomy indeed have different needs and concepts of relevance. The benefits of intent information developed under our taxonomy are applicable to different ranking algorithms.} Especially for some relatively weaker algorithms in this task~(e.g., AdaRank and LambdaMART), the improvements in ranking performance are significant on all evaluation metrics~(t-test, $p<0.05$). Note that we are not intended to propose new ranking models for legal case retrieval in this study. Therefore, we utilized a simple but effective intent-aware adaption framework that could apply to varied ranking models. In conclusion, experimental results suggest the effectiveness of considering the intent taxonomy in the result ranking task. 

\subsection{Summary}
In this section, we integrated the proposed intent taxonomy into two critical downstream IR tasks, i.e., satisfaction prediction and result ranking. In satisfaction prediction, we find that behavioral features play different roles under different search intents. Moreover, involving search intents can improve the performance of satisfaction. In the ranking task, experimental results also suggest the effectiveness of the intent taxonomy by applying an intent-aware adaption. 

\section{Discussions and Implications}
\label{sec: discussion}
Understanding users' search intents is fundamental for search systems to satisfy users' information needs. Towards an in-depth investigation of search intents in legal case retrieval, this paper proposed a novel hierarchical intent taxonomy of legal case retrieval that integrates IR and legal classification theory. 
% To our best knowledge, it is the first taxonomy of user search intents in legal case retrieval. 
% The taxonomy was built iteratively and evaluated extensively via interviews and editorial user studies. 
Regarding \textbf{RQ1}~(\textit{What are the types of user intent in legal case retrieval? }), user search intents can be categorized into five types: search for \textit{Particular Case(s)}, \textit{Characterization}, \textit{Penalty}, \textit{Procedure}, and \textit{Interest}. According to the interviews and editorial studies, the proposed taxonomy has good coverage of the search intents in real-life search practice. Furthermore, the distribution of these intent categories are revealed based on the feedback collected in the verification process. The \textit{Characterization} accounts for the largest proportion among all the intent categories. Meanwhile, the \textit{Penalty} and \textit{Procedure} are also worth research attention. Especially for the \textit{Penalty} intent, all the interviewees have emphasized its importance in legal practice and, meanwhile, the difficulty in satisfying this intent. 

Furthermore, towards modeling user behavior and satisfaction under different search intents~(\textbf{RQ2} and \textbf{RQ3}), we conducted a laboratory user study involving 36 participants majoring in law. Implicit behavioral signals and explicit feedback were collected. Several interesting findings were made. 

Regarding \textbf{RQ2}~(\textit{How \blue{does} user search behavior change with search intents in legal case retrieval? }), we observe significant differences in user behavior under different search intents. In particular, we follow the hierarchical structure of the taxonomy and find differences when applying the criteria successively. Compared to the \textit{Particular Case(s)} intent~(divided by \textbf{Criterion 1}), users tend to be more patient and allocate more time to examine the landing page under the \textit{Learning} intent. \blue{Comparing} the intents classified by \textbf{Criterion 3}~(i.e., \textit{Characterization}, \textit{Penalty}, and \textit{Procedure}), \blue{\textit{Penalty} involves the most effort}. Meanwhile, we observe that users seem quite patient but less satisfied with the clicked results under the \textit{Procedure} intent. 

Regarding \textbf{RQ3}~(\textit{What are the differences in perception and measurement of user satisfaction under different search intents? }), search intents have significant influences on user satisfaction in multiple aspects. We observe that users are less satisfied under the \textit{Learning} intent, especially under the \textit{Procedure}. Although relevance is still the biggest concern in user satisfaction, users care about different factors~(e.g., diversity, region, inspiration, ranking, etc.) given different search intents. With user satisfaction as the ``golden standard'', we find that the popularly adopted online metrics show distinct performance in measuring user satisfaction under different search intents in legal case retrieval. Our results suggest that the optimization and evaluation of search systems may also need to be adapted to different search intents in legal case retrieval. 

Last but not least, we attempted to apply the intent taxonomy to other downstream tasks(e.g., satisfaction prediction and result ranking) to address \textbf{RQ4}~(\textit{How can the taxonomy benefit downstream tasks in legal case retrieval?}). Experimental results demonstrate the benefits of the intent taxonomy in legal case retrieval. 

\paragraph{Implications.} This work provides insight into user intents in the scenario of legal case retrieval. It provides a fundamental research contribution to related studies in legal case retrieval, such as relevance criteria, ranking strategies, and evaluation design. \blue{While our taxonomy was originally developed and validated within the Chinese legal system, it provides a solid foundation that can inspire the development of similar taxonomies in other legal systems. The underlying principles and categorization framework can serve as a starting point for researchers and practitioners working in different jurisdictions.}

Extensive experimental results based on various sources suggest the significance of considering different types of search intents in legal case retrieval. Recent research efforts~\cite{ma2021lecard,shao2021investigating,shao2022understanding} on legal case retrieval are mainly concerned with the \textit{Characterization} tasks, which accounts for the most significant proportion of search intents in legal case retrieval according to our study. However, we argue that other intent types, such as \textit{Penalty} and \textit{Procedure}, are also worth investigation and optimization in the meantime, which still lack due research attention. Moreover, our study has revealed remarkable influences of search intents on various components of an IR system, such as search behavior, user satisfaction, system evaluation, and result ranking. Therefore, our findings suggest that the methodologies in legal case retrieval should be adjusted with various search intents instead of merely similar case matching. This work provides promising implications for the development of legal case retrieval systems to better satisfy users' diverse information needs in practice, such as developing intent-aware ranking strategies and evaluation metrics.

% It provides a fundamental research contribution to further studies, such as relevance criteria, algorithm and evaluation design. Further, this paper illustrates remarkable differences in user search behavior and satisfaction perception under the taxonomy based on a laboratory user study. Our findings suggest that result optimization and system evaluation in legal case retrieval should be adjusted with various search intents instead of merely similar case matching. Last but not least, we constructed intent predictors with the query input as an attempt to aid retrieval system response to different user intents in time. Our work provides promising implications for related research and the development of legal case retrieval systems to better satisfy diverse information needs in practice, such as developing intent-aware ranking strategies and evaluation metrics.
% Each intent category accounts for a non-trivial proportion in realistic legal case retrieval scenarios. 
% 

\section{Conclusion and Future Work}
\label{sec: conclusion}
In this paper, we present a novel intent taxonomy for legal case retrieval. To our best knowledge, it is the first taxonomy that categorizes users' search intents in legal case retrieval. The taxonomy was built based on various resources and further evaluated extensively via interviews and editorial user studies. Furthermore, based on an additional laboratory user study, we discovered significant differences in search behavior and satisfaction under different search intents of legal case retrieval. Finally, we applied the intent taxonomy to two essential tasks in legal case retrieval and demonstrated its implications. 

We acknowledge some potential limitations of this work. The experiments in this paper were mainly conducted based on the Chinese law system, e.g., user studies and query logs, though the taxonomy is designed to be generally applicable across different law systems. The impacts regarding other law systems may need further studying. As for the user study design in Section~\ref{sec:behavior_satis}, the number of participants and tasks is limited as in most user studies, especially for the search scenarios involving domain knowledge~\cite{shao2021investigating,zhang2011predicting}. The \textit{Interest} category still lacks an in-depth inspect limited by the user study environment and data sparsity in query logs. In Section~\ref{sec:application}, traditional models were utilized and the way of integrating search intents seemed straightforward, since our primary concern is the influence of intents on these tasks. More complicated models are out of scope and left for future work. 

As for future work, we will work on developing intent-aware mechanisms for legal case retrieval. For instance, we plan to construct benchmarks with different search intents involved and design intent-aware evaluation metrics. Besides, intent-aware ranking strategies are worth investigating to satisfy diverse information needs better in legal case retrieval. To resolve the user study's limitations, a larger-scale field study will be a promising supplementary for future work. 

% For future work, a field study might be a promising solution. Although the classification theories utilized to construct the taxonomy are applicable in different law systems, the experiments were mainly conducted in the Chinese law system. Its implications in other law systems might need further studying. 
% We acknowledge some potential limitations of this work. The number of users and tasks is limited as in most user studies, especially for the search scenarios involving domain knowledge~\cite{shao2021investigating,zhang2011predicting}. The \textit{Interest} category still lacks an in-depth investigation limited by the user study environment and data sparsity in query logs. For future work, a field study might be a promising solution. Although the classification theories utilized to construct the taxonomy are applicable in different law systems, the experiments were mainly conducted in the Chinese law system. Its implications in other law systems might need further studying. 

\begin{acks}
This work is supported by the Natural Science Foundation of China (Grant No. 61732008, 62002194) and Tsinghua University Guoqiang Research Institute. 
\end{acks}

\newpage
\bibliographystyle{ACM-Reference-Format}
\bibliography{sample-base}

% \appendix

\end{document}